\documentstyle[11pt,newpasp,twoside,amsmath,amssymb,psfig]{article}

\begin{document}

\title{Self-gravitating discs:  what can we learn from the dynamics of maser spots ?}
\author{Jean-Marc Hur\'e}
\affil{LUTh/Observatoire de Paris-Meudon, pl. J. Janssen, F-92195 Meudon}
\affil{ Universit\'e Paris 7, 2 pl. Jussieu, F-75251 Paris Cedex 05}

\setcounter{page}{1}

\begin{abstract}
For a few nearby active nuclei, the disc orbiting the black hole is traced by water maser emission. By combining a simple model together with observed velocity profiles, we show that it is possible to put constraints on the black hole mass and on the distribution of matter (shape, density, size) in the outer disc. We then report possible parameters for the non-keplerian disc and for the black hole in NGC 1068, and mention an uncertainty of at least $25 \%$ on the central mass in NGC 4258.
\end{abstract}

\section{The inverse problem}
Finding the spatial distribution of matter $\rho(\vec{r})$ that reproduces a given rotation curve $\Omega_{\rm obs.}(\vec{r})$ is a common inverse problem in Astrophysics. It can be addressed at the scale of AGN gaseous discs if one interprets the rotation curve of water masers detected in the core of some objects (Greenhill, 2002) in terms of a pure gravitational attraction. Efforts must therefore be made in developing reliable theoretical and/or numerical techniques to perform the inversion (e.g. Sibgatullin, Garcia \& Manko, 2002). In the present situation, this must account for long-range interacting components (black hole, stellar cluster, bars, galactic bludge, etc.) as well as for objects populating the parsec-scale (disc, clouds, compact objects, torus, etc..) where masers are seen. Uncertainties are large, limiting our predictions and interpretations. We discuss here an example of inversion by considering a very simplified, two component system, made of a black hole and a disc, with an application to NGC 4258 and NGC 1068.

\section{Assumptions, equations and method}
At the parsec-scale, the standard theory predicts that all discs are (strongly) self-gravitating (Collin \& Hur\'e, 2000) and so, they might evolve towards a state with a unity $Q$-Toomre parameter. Accounting for disc mass effects is therefore necessary to investigate these outer regions. But a reliable modeling of such discs is not permitted yet, given difficulties of various origins (nature of the flow, turbulent transport, numerical resolution, instabilities, boundary conditions), and some assumptions are necessary. The simplest way to tackle the question is a ``minimal approach'' which assumes that the velocity field has a pure gravitational origin. For a black hole with mass $M_{\rm BH}$ and a disc only, the most basic equation set to solve is
\[\left\{
\begin{split}
&\Omega^2_{\rm obs.} = \frac{GM_{\rm BH}}{r^3} - \frac{g_R^{\rm disc}}{R}\\
&\nabla . {\bf g}^{\rm disc} = 4 \pi G \rho,
\end{split}
\right. \]

where ${\bf g}^{\rm disc}$ is the gravitational acceleration due to the disc, $R$ is the cylindrical radius, $r$ is the spherical radius and $\rho$ is the unknown density, a tri-dimensional function. In other words, if we are able to compute the self-gravitating field for discs with various size and mass distributions, we can select among possible solutions those matching observations. Note that, in this kind of approach, no physical model for the disc imposes the $\rho-$function.

Here, the Poisson equation is solved numerically assuming azimuthal invariance for a given $\rho(\vec{r})$ from the integral form
\begin{equation}
{\bf g}^{\rm disc}(\vec{r}) = - G \iiint_{\rm vol}{\frac{\rho(\vec{r})  ( \vec{r} - \vec{r}' ) d^3 r}{| \vec{r} - \vec{r}' |^3 }},
\end{equation}
by the method described in Hur\'e (2002a). We stress that the full regularization of the volume integral requires three successive integrations, otherwise diverging fields (and subsequently infinite velocities) can result at the disc edges (e.g. Mestel 1963). This can be critical if the maser spots rotate near the disc outer edge. The present inversion method works with three inputs: the black hole mass $M_{\rm BH}$, the mass density in the disc $\rho(R,z)$, and its outer radius $R_{\rm out}$. As several solutions can lead, at least locally, to similar rotation curves, we impose that the total surface density $\Sigma=\int{\rho dz}$ is of the form $\Sigma \propto R^s$, with a smooth connection to the ambient medium at the disc edges (see Hur\'e 2002b for details). The mass density inside the disc obeys a parabolic law with the altitude, and the disc aspect ratio $H/R$ is held constant. These additional restrictions however do not remove the degeneracies. Once a model fits observations, inputs become outputs.

\section{NGC 4258: the possible ambiguity}

\begin{figure}
\psfig{figure=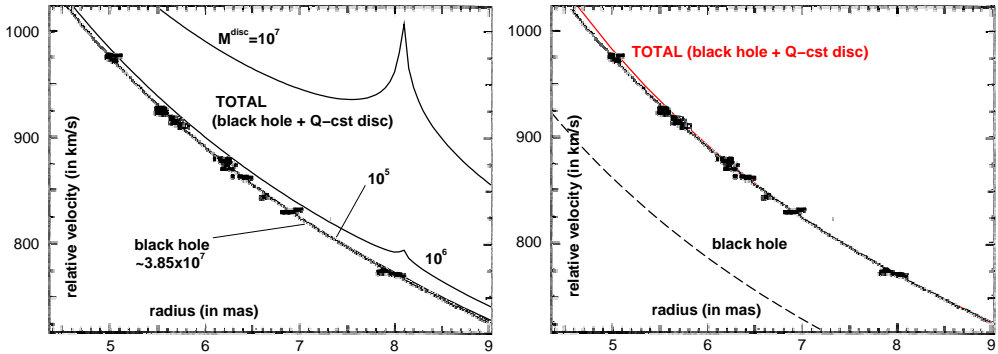,width=13.2cm}
\caption[]{Velocity of masers versus the angular separation in NGC 4258. {\it Left:} theoretical velocity expected for a black hole with nominal mass $M_{\rm BH} \sim 3.9 \times 10^7$ M$_\odot$ and discs with mass of $10^7$, $10^6$ and $10^5$ M$_\odot$ and outer edge at $\sim 8$ mas. {\it Right:} Same but with a $25 \%$ lighter black hole and with a disc with mass $\sim 8 \times 10^6$ M$_\odot$ (see text and Tab. \ref{tab:cmp}).}
\label{fig:NGC4258}
\end{figure}

NGC 4258 is the most famous example of Seyfert-2 active galaxy showing a thin outer disc with masers in quasi-perfect keplerian motion (Myoshi et al. 1995). A priori, this sustains the simple idea that most of the mass is effectively gathered at the center, implying a black hole mass $\sim 3.9 \times 10^7$ M$_\odot$. If this is correct, the inversion method can furnish an upper limit for the disc mass: as figure \ref{fig:NGC4258} shows, the mass of the masing disc at $\sim 0.28$ pc (or $\sim 8$ mas) must be less than $\sim 10^5$ M$_\odot$ (that is a mass ratio $q \lesssim 2.5 \times 10^{-3}$), otherwise a deviation to the keplerian rotation results. But the possibility that a massive disc is present around a lighter black hole must not be ruled out. Actually, the rotation curve concerns a short spatial range and various $R$-laws can mimic a keplerian profile. Figure \ref{fig:NGC4258}b shows a case where the rotation curve is still well reproduced with a black hole $25 \%$ less massive and a disc with a mass $\sim 8 \times 10^6$ M$_\odot$ (or $q\sim 0.26$) and with an outer edge at $0.11$ pc (in this case, the masing clouds rotate ``outside'' the disc). In this example, we have deliberately selected a disc with $Q \sim cst$ (for the reason mentioned before), but the upper limit derived in this way is almost not affected by this specific choice.

\section{NGC 1068}

\begin{figure}
\psfig{figure=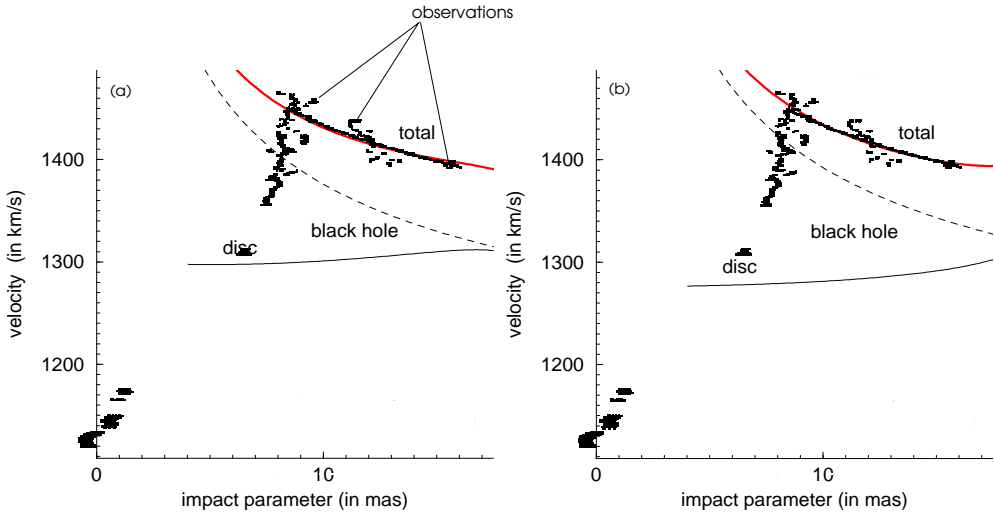,width=13.2cm}
\caption[]{Same as for figure \ref{fig:NGC4258}, but for NGC 1068. {\it Left:} a thick disc solution.  {\it Right:} a thin disc solution (see Tab. \ref{tab:cmp}).}
\label{fig:NGC1068}
\end{figure}

In contrast, the rotation curve of the outer disc in NGC 1068 appears in noticeable departure with respect to the keplerian case. Greenhill et al. (1996) found $\Omega_{\rm obs.} \propto R^{-1.31}$ in the range $0.65 - 1.1$ pc, and mentioned that the disc self-gravity could be the possible source of this non-keplerian behavior. Other mechanisms can be invoked, like radiation pressure effects (Pier \& Krolik 1992) or gravitational attraction by a stellar cluster (Kumar 1999), but the disc self-gravity appears as a natural explanation. Applying the inversion method gives several families of solutions. The sensitivity to $s$ is important. Two possible solutions characterized by a mean angular rotation index of $\sim -1.31$ are shown in Fig. \ref{fig:NGC1068}. The parameters deduced for the disc and central black hole are reported in Table \ref{tab:cmp}. We find that the disc mass is $\gtrsim 10^7$ M$_\odot$, comparable to that of the black hole. We stress that surface densities that do not vary approximately as $R^{-1}$ do not match observations: with $s > -1$, the global rotation curve is flatter (and steeper with $s < -1$). The important point is that the value inferred for the $Q-$parameter is of the order of unity, meaning that the outer disc could be in a marginally stable state. Moreover, the temperature and the density in the outer disc seem appropriate for excitation of water molecules (Hur\'e 2002b).

\begin{table}
\begin{tabular}{|l|cc|cc|} \hline
                   & \multicolumn{2}{|c}{NGC 4258} & \multicolumn{2}{|c|}{NGC 1068}  \\
parameters & light disc &  massive disc & thick disc & thin disc \\ \hline
$M_{\rm bh}$, in M$_\odot$  & $4 \times 10^7$ & $3 \times 10^7$ &   $1.2 \times 10^7$ &  $1.2 \times 10^7$    \\
$R_{\rm out}$, in pc (in mas) & $\sim 0.28$ ($8$) & $\gtrsim 0.11$ ($3.1$) &  $\gtrsim 1.3$ ($18$) & $\gtrsim 1.5$ ($21$)\\
disc aspect ratio $H/R$ & $0.003$ & $0.1$ &  $0.3$ &  $\ll 0.3$ \\
index $s$                      & $-1$ & $-1$ &  $\sim -1.05$ &  $\sim -1$ \\
$q(R_{\rm out})$  & $\le 0.0025$ & $0.26$ & $0.91$ & $0.65$ \\ \hline
\end{tabular}
\caption[]{Possible parameters for the disc and for the central black hole in NGC 4258 and NGC 1068 found from the inversion method.}
\label{tab:cmp}
\end{table}

\section{Conclusions} 

We have discussed possible constraints (and uncertainties) on the mass of the central black hole and on structure of the parsec-scale region in NGC 4258 and NGC 1068, by a method involving a simple axi-symmetrical disc model and a black hole. The subsequent challenges will then be i) to understand, in terms of physical processes, why the disc has the inferred properties, ii) to check the the minimal approach, and iii) to discuss the implications of these solutions for the inner disc (see Lodato \& Bertin in these proceedings).

\acknowledgements
I am grateful to  C. Boisson, S. Collin, S. Courty, F. Hersant, F. Le Petit, 
 D. Richard. and J.-P. Zahn for fruitful discussions.


\begin{references}
\reference Collin S., Hur\'e J.M., 2001, A\&A, 372, 50
\reference Greenhill L.J., Proceedings of IAU Symposium 206: Cosmic Masers (Mangaratiba, Brazil; March 2001)
\reference  Greenhill, L.~J. et al., 1996, \apjl, 472, L21 
\reference Hur\'e J.M., 2002a, submitted
\reference Hur\'e J.M., 2002b, A\&A Let., in press
\reference Kumar P., 1999, ApJ, {\bf 519}, 599
\reference Mestel L., 1963, ApJ, {\bf 126}, 553
\reference Miyoshi, M. et al. 1995, Nature, {\bf 373}, 127
\reference Pier, E.~A., Krolik J.~H.~, 1992, ApJL, {\bf 399}, L23-L26.
\reference Sibgatullin, N.R., Garcia, G., Manko, V.S., 2002, astro-ph/0209200
\end{references}
\end{document}